# FITTING ASTRONOMICAL DATA

André Le Floch

University of Tours, Department of Physics

37200 – Tours, France

**Abstract**

Deming's method is applied for calculating matrix elements allowing to fit orbital parameters for planets. This work provides demonstrations which were missing in our previous paper of 2002.

## Introduction

Matrix elements allowing to fit the orbital parameters of planets have been collected and used in our preceding work (2002, called paper 1). However, these matrix elements were listed without any demonstration. So, the aim of the present work is to provide these missing demonstrations.

## 1. Symbols

First, in the case of measurements of planet positions, we are dealing with three variables (time, right ascension, and declination) connected with $q$ objects (a planet, a satellite or the Sun) . Thus, for each object $l$ ( with $l = 1,2,\ldots,q$ and $i_l = 1$ to $n$) where $n$ is the total number of observations, $ti_l, \alpha i_l, \delta i_l$ and $\overline{ti_l}, \overline{\alpha i_l}, \overline{\delta i_l}$ represent the observed and calculated quantities, respectively.

Their statistical weights are given by:

$$Wti_l = \frac{\sigma_o^2}{(\sigma(ti_l))^2} \; ; \quad W\alpha i_l = \frac{\sigma_o^2}{(\sigma(\alpha i_l))^2} \; ; \quad W\delta i_l = \frac{\sigma_o^2}{(\sigma(\delta i_l))^2} \qquad \text{(eq. 1.1)}$$

where $\sigma(ti_l), \sigma(\alpha i_l), \sigma(\delta i_l)$ are the corresponding uncertainties, and $\sigma_o^2$ the arbitrary variance of unit weight (see Deming’s textbook).

The residuals are (observed - calculated):

$$Vti_l = ti_l - \overline{ti_l} \quad ; \quad V\alpha i_l = \alpha i_l - \overline{\alpha i_l} \quad ; \quad V\delta i_l = \delta i_l - \overline{\delta i_l} \qquad \text{(eq. 1.2)}$$

## 2 . The condition equations

Following the original and general Deming’s method, we shall suppose that the adjusted values $\overline{ti_l}, \overline{\alpha i_l}, \overline{\delta i_l}$ of the three variables, and the adjusted values of the unknown $p$ parameters $\overline{a_1}, \overline{a_2},\ldots,\overline{a_j},\ldots,\overline{a_p}$ , are connected by the following condition equations:

$$\overset{(l)}{A}\,(\overline{\alpha i_l}, \overline{ti_l}, \overline{a_1}, \overline{a_2},\ldots,\overline{a_j},\ldots,\overline{a_p}) = 0 \qquad \text{(eq. 2.1)}$$

$$\overset{(l)}{D}\,(\overline{\delta i_l}, \overline{ti_l}, \overline{a_1}, \overline{a_2},\ldots,\overline{a_j},\ldots,\overline{a_p}) = 0 \qquad \text{(eq. 2.2)}$$

We also suppose that functions $\overset{(l)}{A}$ and $\overset{(l)}{D}$ are linearly dependent of variables $\overline{\alpha i_l}$ and $\overline{\delta i_l}$, respectively. In other words we suppose that:

$$\partial \overset{(l)}{A} \,/\, \partial \overline{\alpha i_l} = 1 \quad \text{and} \quad \partial \overset{(l)}{D} \,/\, \partial \overline{\delta i_l} = 1 \qquad \text{(eq. 2.3)}$$

Thus, we have $2q$ condition equations involving the estimates $\overline{a_1}, \overline{a_2},\ldots,\overline{a_j},\ldots,\overline{a_p}$ of the unknown $p$ parameters, and the calculated values of the variables $\overline{ti_l}, \overline{\alpha i_l}, \overline{\delta i_l}$ .

We then write:

$$\overset{0}{A}i_l = \overset{(l)}{A}\,(\alpha i_l, ti_l, \overset{0}{a_1}, \overset{0}{a_2},\ldots,\overset{0}{a_j},\ldots,\overset{0}{a_p})$$

$$\overset{0}{Di_l} = \overset{(l)}{D} (\delta i_l , ti_l, \overset{0}{a_1}, \overset{0}{a_2} ,\ldots,\overset{0}{a_j},\ldots,\overset{0}{a_p} )$$

where $\overset{0}{Ai_l}$ and $\overset{0}{Di_l}$ are just the amount by which the condition equations fail to be satisfied by the observed values and the initial approximations $\overset{0}{a_1}, \overset{0}{a_2} ,\ldots,\overset{0}{a_j},\ldots,\overset{0}{a_p}$ of parameters.

## 3 . Taylor expansions

Expanding condition equations by Taylor's series to first order of residuals, we obtain:

$$\overset{0}{Ai_l} = V\alpha i_l + Vti_l \, (\partial \overset{(l)}{A} / \partial ti_l) + \sum_{J=1}^{p} ( \partial \overset{(l)}{A} / \partial a_j ) \Delta a_j \qquad \text{(eq. 3.1)}$$

$$\overset{0}{Di_l} = V\delta i_l + Vti_l \, (\partial \overset{(l)}{D} / \partial ti_l) + \sum_{k=1}^{p} ( \partial \overset{(l)}{D} / \partial a_k) \Delta a_k \qquad \text{(eq. 3.2)}$$

with

$$a_j = \overset{0}{a_j} - \Delta a_j \qquad ( j = 1,\ldots.,p)$$

The variational conditions for these reduced conditions:

$$d ( \overset{0}{Ai_l} ) = 0 \quad \text{and} \quad d(\overset{0}{Di_l}) = 0$$

gives:

$$0 = d (V\alpha i_l ) + ( \partial \overset{(l)}{A} / \partial ti_l) \, d (Vti_l ) + \sum_{J=1}^{p} ( \partial \overset{(l)}{A} / \partial a_j ) \, d(\Delta a_j)$$

$$0 = d (V\delta i_l ) + ( \partial \overset{(l)}{D} / \partial ti_l) \, d (Vti_l ) + \sum_{k=1}^{p} ( \partial \overset{(l)}{D} / \partial a_k) \, d(\Delta a_k)$$

for $l = 1$ to $q$ and $i_l = 1$ to $n$ (eqs. 3.3 and 3.4)

## 4 . The least squares condition

The sum of the weighted squares of the residuals will be minimum:

$$S = \sum_{i_l=1}^{n} Si_l = minimum$$

with (eq. 4.1)

$$Si_l = Wti_l\,(Vti_l)^2 + W\alpha i_l(V\alpha i_l)^2 + W\delta i_l(V\delta i_l)^2$$

The minimum condition $dS = O$ gives:

$$O = \sum_{i_l}\left(Wti_l\, Vti_l\, d(Vti_l) + W\alpha i_l\, V\alpha i_l\, d(V\alpha i_l) + W\delta i_l\, V\delta i_l\, d(V\delta i_l)\right)$$

(eq. 4.2)

## 5 . The method of Lagrange multipliers

As suggested by Deming, minimization of the least squares condition (eq. 4.1) could be solved by introducing $2\,n$ arbitrary *Lagrange multipliers* :

$-\lambda i_l$ in eq. (3.3) and $-\mu i_l$ in eq. (3.4) .

Then, by adding eqs. (3.3)+(3.4)+(4.2), we get a linear expression of differentials equal to zero.

Thus, all coefficients of these differentials must be equal to zero:

$$W\alpha i_l\, V\alpha i_l - \lambda i_l = O \qquad \text{(eq. 5.1)}$$

$$W\delta i_l\, V\delta i_l - \mu i_l = O \qquad \text{(eq. 5.2)}$$

$$Wti_l\, Vti_l - \lambda i_l\,(\partial A^{(l)} / \partial ti_l) - \mu i_l\,(\partial D^{(l)} / \partial ti_l) = O \qquad \text{(eq. 5.3)}$$

$$\sum_{i_l=1}^{n}\left(\lambda i_l\,(\partial A^{(l)} / \partial a_j) + \mu i_l\,(\partial D^{(l)} / \partial a_j)\right) = O \qquad \text{(eq. 5.4)}$$

( for $j = 1$ to $p$ )

Thus, we have obtained $5\,n + p$ equations (eqs. 3.1, 3.2 and eqs. 5) for the $5\,n + p$ unknowns: $V\alpha i_l$ , $V\delta i_l$ , $Vti_l$ , $\lambda i_l$ , $\mu i_l$ , $\Delta a_j$ . In other words, this system is mathematically determinate, and the solution can be obtained by substitutions.

## 6 . Corrections to parameters

Now, our next objective is to obtain a matrix relation allowing to calculate the correction vector $\Delta a_j$ to parameters. This vector could be obtained by resolving the system of equations 3 and 5. In a first step we shall extract $V\alpha i_l$ , $V\delta i_l$ , $Vti_l$ from eqs. (5.1), (5.2), (5.3), respectively. Thus we have:

$$V\alpha i_l = \lambda i_l / W\alpha i_l \ , \ V\delta i_l = \mu i_l / W\delta i_l \qquad \text{(eqs. 6.1 and 6.2)}$$

$$Vti_l = \left( \lambda i_l \left( \partial \overset{(l)}{A} / \partial ti_l \right) + \mu i_l \left( \partial \overset{(l)}{D} / \partial ti_l \right) \right) / Wti_l \qquad \text{(eq. 6.3)}$$

Then, we introduce preceding values of $V\alpha i_l$ , $V\delta i_l$ , $Vti_l$ in (eqs. 3.1 and 3.2)

Thus, from (eq. 3.1) we may obtain $\lambda i_l$ as a function of $\overset{0}{A} i_l$ and $\mu i_l$ (with several coefficients). The next step consists to transport this expression of $\lambda i_l$ in (eq. 3.2) . So, we obtain (with symbols collected in the Appendix) the required values of $\lambda i_l$ and $\mu i_l$ :

$$\lambda i_l = s i_l \left( q i_l \, b i_l - r i_l \, d i_l \right) \qquad \text{(eq. 6.4)}$$

and

$$\mu i_l = s i_l \left( p i_l \, d i_l - r i_l \, b i_l \right) \qquad \text{(eq. 6.5)}$$

Then eq. (5.4) gives :

$$\sum_{i_l=1}^{n} \left( s i_l (q i_l \, b i_l - r i_l \, d i_l)\left( \partial \overset{(l)}{A} / \partial a_j \right) + s i_l (p i_l \, d i_l - r i_l \, b i_l)\left( \partial \overset{(l)}{D} / \partial a_j \right) \right) = 0$$

$$( \text{for } j = 1 \text{ to } p ) \qquad \text{(eq. 6.6)}$$

It is also useful to define the error-vector $\Delta W$ whose elements are:

$$\Delta W(j) = \sum_{i_l=1}^{n} \left( u i_l(j) \overset{0}{A} i_l + v i_l(j) \overset{0}{D} i_l \right) \qquad \text{(eq. 6.7)}$$

$$( \text{for } j = 1 \text{ to } p )$$

Consequently, (eq. 6.6) could be written as :

$$\sum_{i_l=1}^{n}\left( v_{i_l}(j)\sum_{k=1}^{P}(\partial D^{(l)}/\partial a_k)\Delta a_k + u_{i_l}(j)\sum_{k=1}^{p}(\partial A^{(l)}/\partial a_k)\Delta a_k \right) = \Delta W(j)$$

( for $j = 1$ to $p$ ) (eq. 6.8)

and also as a matrix equation:

$$N\,\Delta A = \Delta W \quad \text{equivalent to} \quad \Delta A = N^{-1}\,\Delta W$$ (eq. 6.9)

where $\Delta A$ is the correction-vector (with components $\Delta a_k$) and $N$ the normal equation matrix (square-symmetric of dimensions $p \times p$ ) whose elements are given by:

$$N(j,k) = \sum_{i_l=1}^{n}\left( u_{i_l}(j)\times(\partial A^{(l)}/\partial a_k) + v_{i_l}(j)\times(\partial D^{(l)}/\partial a_k)\right)$$

(eq. 6.10)

For numerical calculations, we now remember the linear relationships (eqs. 2.3) between functions $A^{(l)}$ and $D^{(l)}$ with $\bar{\alpha}_{i_l}$ and $\bar{\delta}_{i_l}$, respectively. So, various derivatives appearing in preceding relations could be calculated by replacing $A^{(l)}$ by $\bar{\alpha}_{i_l}$ and $D^{(l)}$ by $\bar{\delta}_{i_l}$ in the relevant derivatives. It is also the same for the initial residuals of right ascension and declination, we have:

$$A^0_{i_l} = V\alpha^0_{i_l} \quad \text{and} \quad D^0_{i_l} = V\delta^0_{i_l}$$

So, we can calculate (numerically) the fundamental equations (6.7) and (6.10) and the correction-vector (eq.6.9) to the initial approximate parameters.

Thus we get back the same expressions as in paper 1 (after correcting two errors appearing in the Appendix A of paper 1, corresponding to eqs. (6.3) and (6.5) of this work).

## 7 . The variance – covariance matrix

First, remember that the *goodness* of the fit is given by the *reduced variance* $\sigma$ defined by (see eq. 4.1):

$$\sigma = \left( S / (n\text{-}p) \right)^{1/2} = \left( \left( \sum_{i_l=1}^{n} S i_l \right) / (n\text{-}p) \right)^{1/2}$$

Because we have three variables, it is convenient to put the arbitrary coefficient

$\sigma_o^2 = 1/3$. Thus, as $n >> p$ we remark that:

$$perfect\ fit \Leftrightarrow \sigma \cong 1$$

Now, the variance – covariance matrix $V$ could be defined as follows:

$$V(j,k) = \sigma^2 N^{-1}(j,k) \qquad (\text{ for } j,k = 1 \text{ to } p)$$

It is important to remark that the arbitrary coefficient $\sigma_o$ disappears in the expression of $V$ (it appears both at the numerator and denominator of each matrix element).

Recall that *diagonal elements* of $V$ are the squares of the *standard errors* of the fitted parameters. Furthermore, *off-diagonal elements* of $V$ provides the *correlation coefficients* $c(j,k)$ between the fitted parameters. We have:

$$c(j,k) = V(j,k) / \left( V(j,j) \times V(k,k) \right)^{1/2} \qquad (\text{ for } j,k = 1 \text{ to } p)$$

As we have shown in paper 1, a remarkable example is provided by the Solar mass and the Gravitational constant, which are very highly correlated.

More generally, if $f(a_1, a_2,\ldots,a_j,\ldots,a_p)$ is a function of the fitted parameters, the square of the standard error propagated on $f$ is:

$$\sigma(f)^2 = \sum_{j=1}^{p} \sum_{k=1}^{p} (\partial f / \partial a_j)(\partial f / \partial a_k) V(j,k)$$

For example, this expression should be used for *long-time extrapolations* of astronomical quantities. So, when the error propagated on $f$ is becoming of the order of this quantity, those extrapolations could be meaningless. This is, for instance, the case of long-time extrapolations dealing with the semi-major axis of the Earth orbit.

## 8 . Conclusion

Thus we can conclude that this work provides the missing demonstrations of our previous paper of 2002 (with corrections of two errors).

## Appendix

It is convenient (in preceding calculations, and for coding programs) to use the following symbols (similar to those listed in paper 1):

$$pi_l = (1 / W\alpha i_l) + (\partial A^{(l)} / \partial ti_l)^2 / Wti_l$$

$$qi_l = (1 / W\delta i_l) + (\partial D^{(l)} / \partial ti_l)^2 / Wti_l$$

$$ri_l = \left( (\partial A^{(l)} / \partial ti_l)(\partial D^{(l)} / \partial ti_l) \right) / Wti_l$$

$$si_l = 1 / (pi_l \, qi_l - ri_l^{\,2})$$

$$ui_l(j) = si_l \left( qi_l (\partial A^{(l)} / \partial a_j) - ri_l (\partial D^{(l)} / \partial a_j) \right)$$

$$vi_l(j) = si_l \left( pi_l (\partial D^{(l)} / \partial a_j) - ri_l (\partial A^{(l)} / \partial a_j) \right)$$

$$bi_l = A^0 i_l - \sum_{k=1}^{p} \left( (\partial A^{(l)} / \partial a_k) \Delta a_k \right)$$

$$di_l = D^0 i_l - \sum_{k=1}^{p} \left( (\partial D^{(l)} / \partial a_k) \Delta a_k \right)$$

## Acknowledgements

The author is greatly indebted to Dr. Manuel Sanchez for his invitation to present this work (in May 2002) at San Fernando Observatory, Andalucia (Spain).